\documentclass[pdflatex,sn-nature]{sn-jnl}%

\usepackage{graphicx}%
\usepackage{multirow}%
\usepackage{amsmath,amssymb,amsfonts}%
\usepackage{amsthm}%
\usepackage{mathrsfs}%
\usepackage[title]{appendix}%
\usepackage{xcolor}%
\usepackage{textcomp}%
\usepackage{manyfoot}%
\usepackage{booktabs}%
\usepackage{algorithm}%
\usepackage{algorithmicx}%
\usepackage{algpseudocode}%
\usepackage{listings}%

\usepackage{xurl} %

\theoremstyle{thmstyleone}%

\theoremstyle{thmstyletwo}%

\theoremstyle{thmstylethree}%

\begin{document}

\title[Article Title]{Dialogue Design Shapes the Intergroup Attitudinal Effects of AI-Assisted Political Argumentation}

\author*[1]{\fnm{Jianlong} \sur{Zhu}}\email{jzhu@cs.uni-saarland.de}

\author[1]{\fnm{Syed Muhammad Jhon Raza} \sur{Naqvi}}%

\author[2]{\fnm{Carolin-Theresa} \sur{Ziemer}}%

\author[3]{\fnm{Usman} \sur{Naseem}}%

\author[1]{\fnm{Ingmar} \sur{Weber}}%

\affil*[1]{\orgdiv{Department of Computer Science}, \orgname{Saarland University}, \orgaddress{\street{Saarland Informatics Campus E1.7}, \city{Saarbrücken}, \postcode{66123}, \state{Saarland}, \country{Germany}}}
\affil[2]{ \orgname{Max Planck Institute for Human Development}, \orgaddress{\street{Lentzallee 94}, \postcode{14195}, \state{Berlin}, \country{Germany}}}
\affil[3]{\orgdiv{School of Computing}, \orgname{Macquarie University}, \orgaddress{\street{4 Research Park Dr}, \city{Sydney}, \postcode{2113}, \state{New South Wales}, \country{Australia}}}

\abstract{
Structured argumentative dialogues where interlocutors deliberate on opposing political ideas are known to promote perspective-taking and reduce political polarization, but finding willing partners is difficult as Americans increasingly shun political discussions. AI dialogue partners offer a scalable framework for such open-mindedness exercises, but how the format of human-AI dialogues shapes their benefits remains unclear. This study seeks to fill the gap with a preregistered two-session online experiment with 527 US participants. As the primary experimental manipulation, participants were assigned to argue either \textit{for} or \textit{against} their pre-existing attitude on a contested political issue, engaging either with an AI chatbot or a solitary essay task. The AI conditions further varied in the chatbot’s interaction style (adversarial or collaborative) and the presence of an additional financial incentive. The results show that attitude-congruent dialogues more strongly reduced polarization than attitude-incongruent dialogues immediately after the exchange. By contrast, an exploratory analysis suggested a delayed increase in cognitive empathy following attitude-incongruent dialogues, a pattern consistent with the account of sleeper effects. While the conversation style had little influence on the effects of attitude-congruent dialogues, a collaborative discussion tended to make attitude-incongruent dialogues more effective, narrowing the immediate effect gap. Additional financial incentives did not alter outcomes. Given the heterogeneity, the AI conditions were not universally more effective forming favorable intergroup attitudes in pooled comparisons between AI and non-AI conditions. The findings caution against a simplistic view of AI dialogues as a silver bullet for depolarization and highlight dialogue design as a key determinant of effective AI-mediated attitudinal interventions.
}


\maketitle

\section*{}

In a highly polarized society, citizens not only hold seemingly irreconcilable opinions but also form negative perceptions of out-partisans \cite{iyengar_affect_2012, pasek_misperceptions_2022}, which can result in reduced intergroup contact that further deepens the political divide. The disconnection calls for a structured approach to engineer productive exchanges to build mutual understanding.

Argumentative dialogues that expose interlocutors to cross-cutting views and encourage perspective-taking are commonly seen as a beneficial exercise for a democratic society \cite{mutz_hearing_2006, schneider_need_2023, caluwaerts_deliberation_2023, fishkin_is_2021}. Competitive debate, a form of structured argumentation in which participants argue assigned positions in a small group, has been shown to be effective in achieving certain educational and intergroup outcomes, such as enhanced knowledge\cite{kennedy_power_2009}, deeper critical thinking \cite{chen_exploring_2022}, reduced aggression in conflict management \cite{de_conti_impact_2014}, and reduced confirmation bias and opinion polarization over controversial issues \cite{budesheim_consider_1999, de_conti_debate_2013}.

Despite the promising evidence from argumentative dialogues in a controlled environment, overly negative perceptions \cite{pasek_misperceptions_2022} of out-partisans and fear of social discomfort from disagreement \cite{gerber_disagreement_2012, eddy_americans_2026, boland_zero-sum_2024} still present significant social barriers for such exercises to take place at scale. 
One approach to overcome such barriers is anonymous online cross-partisan messaging, supported by evidence that people feel less restrained when communicating privately online \cite{suler_online_2004}. 
Its effectiveness in reducing polarization, however, is contingent on the civility of willing out-partisans and the mildness of disagreement \cite{combs_reducing_2023, de_jong_cross-partisan_2024}. 
AI chatbots, which have become popular as an everyday conversation partner for information-seeking, offer an alternative route to bypass the interpersonal risks. Studies show people are less fearful of judgment when conversing with an AI chatbot \cite{croes_digital_2024} and are more receptive to counter-attitudinal messages coming from an AI source \cite{lu_how_2025, willis_bridging_2026, hruschka_reducing_2026}. Furthermore, AI models are shown to be capable of inducing lasting attitude change
\cite{costello_durably_2024, sevi_ai_2026}.

While prior research has explored ways to configure AI models as an argumentative partner \cite{farag_opening_2022, durmus_exploring_2018, slonim_autonomous_2021} and presented evidence of AI chatbots' potential to reduce polarization through active listening \cite{hruschka_reducing_2026} and persuasion \cite{walter_using_2025}, less attention has been paid to empirically testing whether AI can reduce polarization by actively engaging users in a dialogue format intended to promote perspective-taking and self-persuasion.

In this study, we explored human-AI interaction as a scalable framework to deliver the societal benefits of argumentative dialogues to the US populace, which has become increasingly divided along the lines of the two main political camps \cite{eddy_americans_2026}. In particular, we sought evidence on whether an AI partner can depolarize more effectively by supporting interlocutors in their active counterargument generation or by directly providing counterarguments as a pleasant but persuasive out-partisan, and examined factors that can potentially enhance engagement. 
To test competing theories of argumentation in the context of human-AI interaction, the study explored dialogue design with three experimental variables: (i) argument side, (ii) dialogue mode, and (iii) financial incentive.

First, the argument side was defined  by the relation between an interlocutor's own attitude and the position they are required to take. We describe participants who made arguments that were congruent with their own attitudes as engaging in pro-attitudinal advocacy. Those who played ``devil's advocate", or argued for a position they disagreed with, performed counter-attitudinal advocacy\footnote{Hereafter we use the terms ``attitude-congruent dialogue” and ``pro-attitudinal dialogue” interchangeably to refer to exchanges in which participants were asked to advance arguments consistent with their own views. Conversely, ``attitude-incongruent dialogue” and ``counter-attitudinal dialogue” are used interchangeably to refer to conditions in which participants were required to argue against their own views.}. Existing literature on debate as a classroom-based exercise largely supports assigning students to argue against their true position to avoid entrenchment, as experimental results showed reduced opinion polarization \cite{budesheim_consider_1999, de_conti_debate_2013} for students who played devil's advocate. In theories of persuasion, counter-attitudinal advocacy prompts self-persuasion, which creates stronger and more durable attitude change than receiving arguments generated by others \cite{miller_counter-attitudinal_2001, xu_persuasive_2024, weiner_persuasion_2003}. 
Evidence also suggests, however, that the advantage of self-generated arguments in persuasion diminishes as the difficulty of argument generation rises \cite{xu_persuasive_2024}, making it worthwhile to investigate whether counter-attitudinal advocacy is still preferable in human-AI dialogue.

Second, the dialogue mode was defined at two levels, adversarial debate and collaborative discussion, in recognition that interlocutors can exchange their opposing views in different tones and dynamics. Debate, with its apparent goal of winning, encourages interlocutors to think out their arguments \cite{gurin_dialogue_2013}. While debate has been shown to be an effective form of classroom pedagogy, the exchange of sharp arguments in an adversarial manner may not always bring the two sides closer in an uncontrolled environment. Unlike offline intergroup interaction \cite{hodson_ideologically_2011}, cross-partisan exchange online can easily turn tribal and uncivil, often increasing political polarization \cite{bliuc_online_2021, bail_exposure_2018}. A less confrontational discussion, in contrast, positions issue understanding as a goal and encourages interlocutors to absorb---as opposed to refute---information \cite{gurin_dialogue_2013, bosser_students_2020}. A discussion can also frame the two sides as collaborators working towards common ground, which enables the possibility to reduce political polarization through a negotiated resolution: treating an agreement as a win for both sides rather than a defeat in a zero-sum contest \cite{iwry_open_2021}. 
In the context of human-AI interaction, there has been limited empirical evidence on whether AI's conversational receptiveness through a collaborative discussion can achieve greater intergroup attitude changes than firm disputation in a debate. Furthermore, relating to the design of an argumentative dialogue, it is worth investigating whether the dialogue modes' contrasting potentials in enhancing articulation and receptiveness result in different effects across argument side.

Third, a financial incentive was randomly presented as a small performance-based bonus on top of our advertised compensation rate for survey completion. For those who were required to play the difficult role of counter-attitudinal advocacy in an argumentative dialogue, the inclusion of the experimental variable was intended to test whether financial incentive could provide interlocutors with additional motivation to ``steelman" the other side instead of giving up. The existing evidence is two-sided. On one side, financial incentive has been shown to increase task compliance \cite{miller_counter-attitudinal_2001} and reduce motivated reasoning in the quest to change attitudes or beliefs \cite{linder_decision_1967, ziemer_psychological_2024, rathje_accuracy_2023, ronzani_how_2024}. On the other, studies also show monetary reward can crowd out intrinsic motivation \cite{murayama_neural_2010, vlaev_changing_2019}, weakening the effects that hinge on deep voluntary engagement.

Through a preregistered between-subjects online experiment (N=527; 469 across AI conditions, 58 in non-AI reference conditions), we examined the effects of eight treatment conditions of human-AI political dialogue on four outcome areas relating to intergroup attitudes: affective polarization, cognitive empathy, support for democratic norms and political intolerance. US participants recruited from Prolific with census-matched demographic quota were assigned to either a non‑AI writing reference task (pro‑ or counter‑attitudinal essay) or an AI‑dialogue arm with a 2×2×2 factorial design crossing argument side (pro‑ or counter‑attitudinal advocacy), dialogue mode (collaborative discussion or adversarial debate), and financial incentive (absent or present). Participants who completed Session 1 and passed our preregistered quality checks were invited to return 14 days later for Session 2, in which they were assigned the same task and experimental condition, but with a different discourse topic. The study setup is visualized in Figure \ref{fig_study_setup} and further details on participant recruitment can be found in Supplementary Section 1.1.

\begin{figure*}[t] %
\centering
\includegraphics[width=0.9\textwidth]{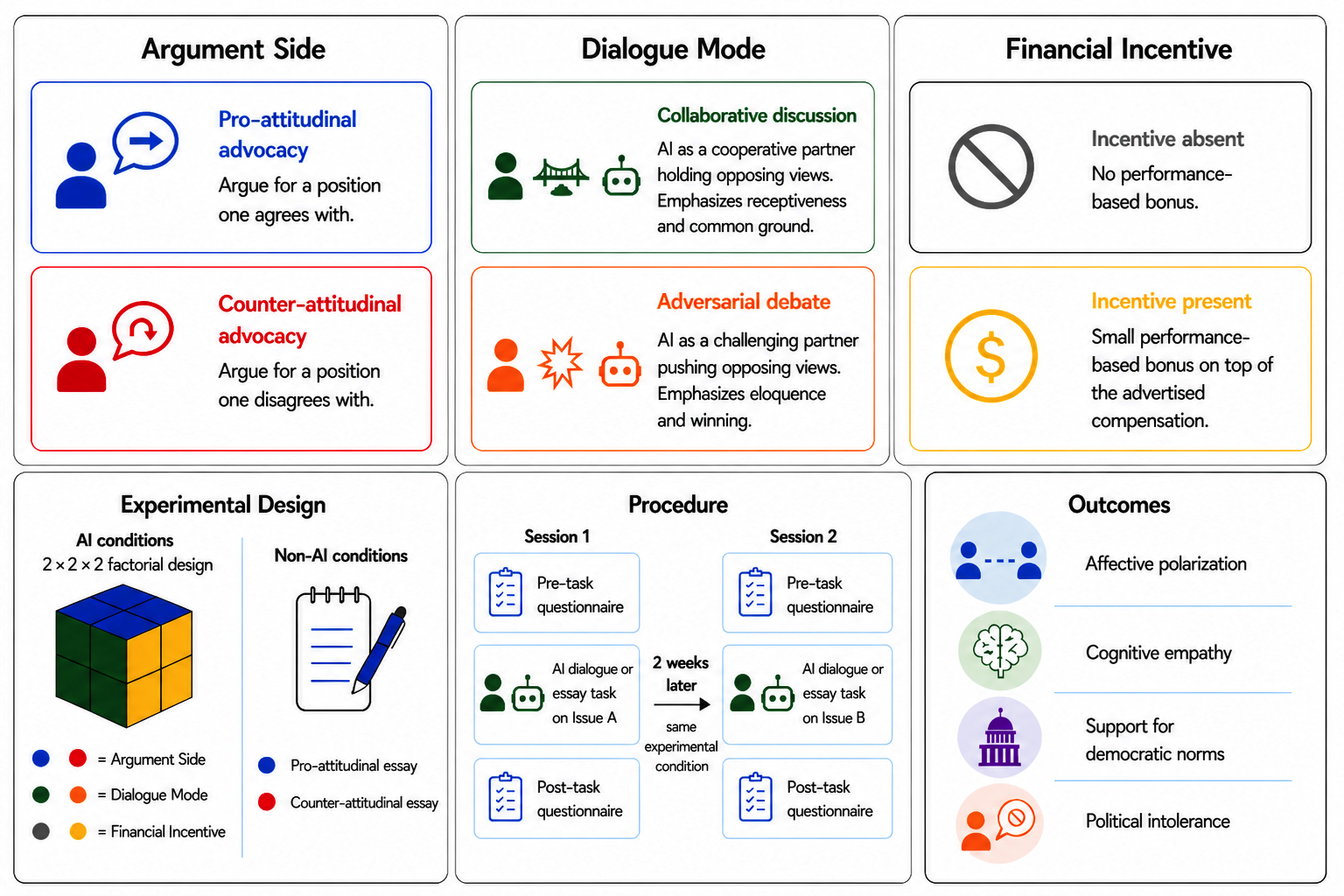}
\caption{Overview of the study's experimental design, procedure and outcome variables.}
\label{fig_study_setup}
\end{figure*}

\begin{figure*}[t]
\centering
\includegraphics[width=1.0\textwidth]{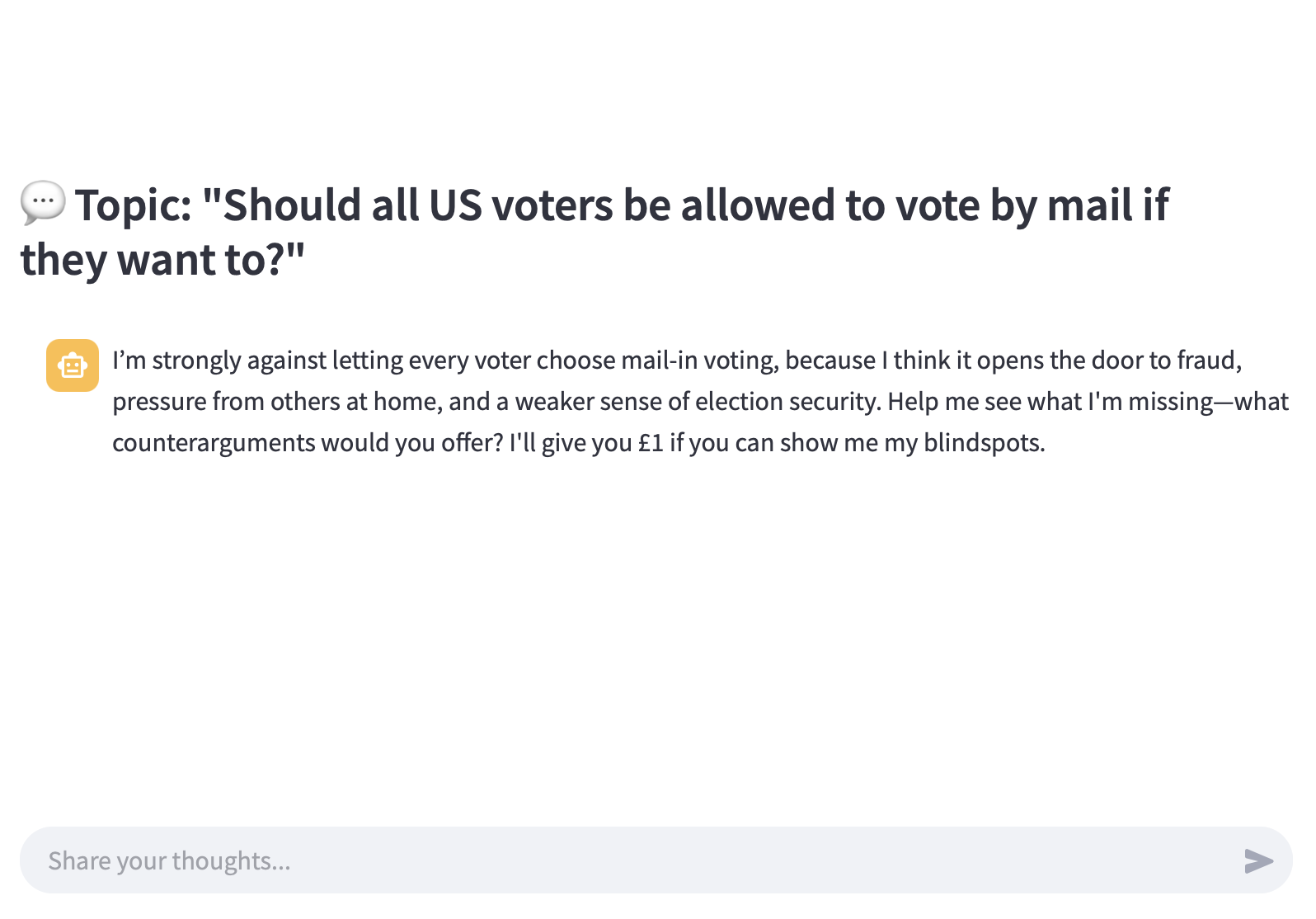}
\caption{An example opening message from the AI chatbot, where the dialogue mode is collaborative discussion and an additional financial incentive is present.}\label{fig_screenshot}
\end{figure*}

Prospective participants with profiles matching our inclusion requirements on Prolific were shown a task description, which broadly outlined the study as data collection for AI training. To minimize demand effect, the goal of measuring attitude change was only disclosed in a debriefing message at the conclusion of the study. Those who accepted the task were taken to a web survey, which included either a chatbot (Figure \ref{fig_screenshot}) or text input fields. Based on survey responses, the participants were shown one of four possible issue statements\footnote{Session 1 covered the topics of postal voting, CDC's health policies, unionization, and oil and gas drilling. Supreme Court expansion, the Ukraine war, Trump's economic policies, and DEI practices were mentioned in Session 2.} and an opening message from the AI dialogue partner, to which they were instructed to construct counterarguments.

We found that the effects of AI intervention varied greatly across conditions, chiefly between the two different argument side values. 
The pro-attitudinal dialogues, which made the AI partner responsible for presenting opposing viewpoints, resulted in greater immediate reduction in both affective ($d=0.21$; Figure \ref{fig_dotplot_affective_differentiation}) and opinion ($d=0.20$) polarization relative to the counter-attitudinal dialogues, as measured by opinions of people who agree or disagree with a political issue and by stances on the issue itself. Whereas the pro-attitudinal AI dialogues showed a marginally significant advantage over the pro-attitudinal non-AI reference task in the reduction of affective polarization ($p=0.058, d=0.33$), arguing counter-attitudinally with an AI showed no greater benefits than the corresponding reference task ($p=0.681, d=0.08$).
Contrary to evidence from classroom settings, playing devil's advocate against an AI partner could temporarily detract from a perspective-taking exercise, resulting in lower levels of cognitive state empathy ($d=0.53$)---understanding of the perspectives of ``the other side" in the moment---than the corresponding non-AI counter-attitudinal essay task. In a reversal of its short-term underperformance, performing counter-attitudinal advocacy against the AI partner increased cognitive trait empathy---which is stable over time and less sensitive to changes in the state of mind---in the two-week period following Session 1 ($d=0.17$).

The results suggest that the AI partner's ability to show alternative perspectives and shift opinions drove immediate benefits in an attitude-congruent dialogue. By contrast, the benefits of the deliberative task of generating attitude-incongruent arguments surfaced later on.

Our analysis presents a nuanced picture on the suitability of an AI-based argumentative dialogue to improve intergroup relations.
While promising as a scalable intervention, the presence of AI alone may not lock in additional benefits for an argumentative dialogue, since success is influenced by argument side and time horizon. For civic stakeholders looking to explore and validate AI-mediated pedagogy for argumentation, the implication is that existing expectations about classroom-based argumentative exercises may not be directly transferable to the context of human-AI interaction. The two argument sides appear to operate through distinct mechanisms and produce heterogeneous outcomes, hence intentional design choices should be made to suit the teaching plan and the needs of students. Above all, dialogue design is a key determinant of success for an argumentative dialogue with an AI chatbot, with implications for AI-mediated behavioral interventions more broadly.

\section*{Results}

\subsection*{Attitude-congruent AI dialogues lead to greater immediate depolarization}

\begin{figure*}[t]
\centering
\includegraphics[width=1.0\textwidth]{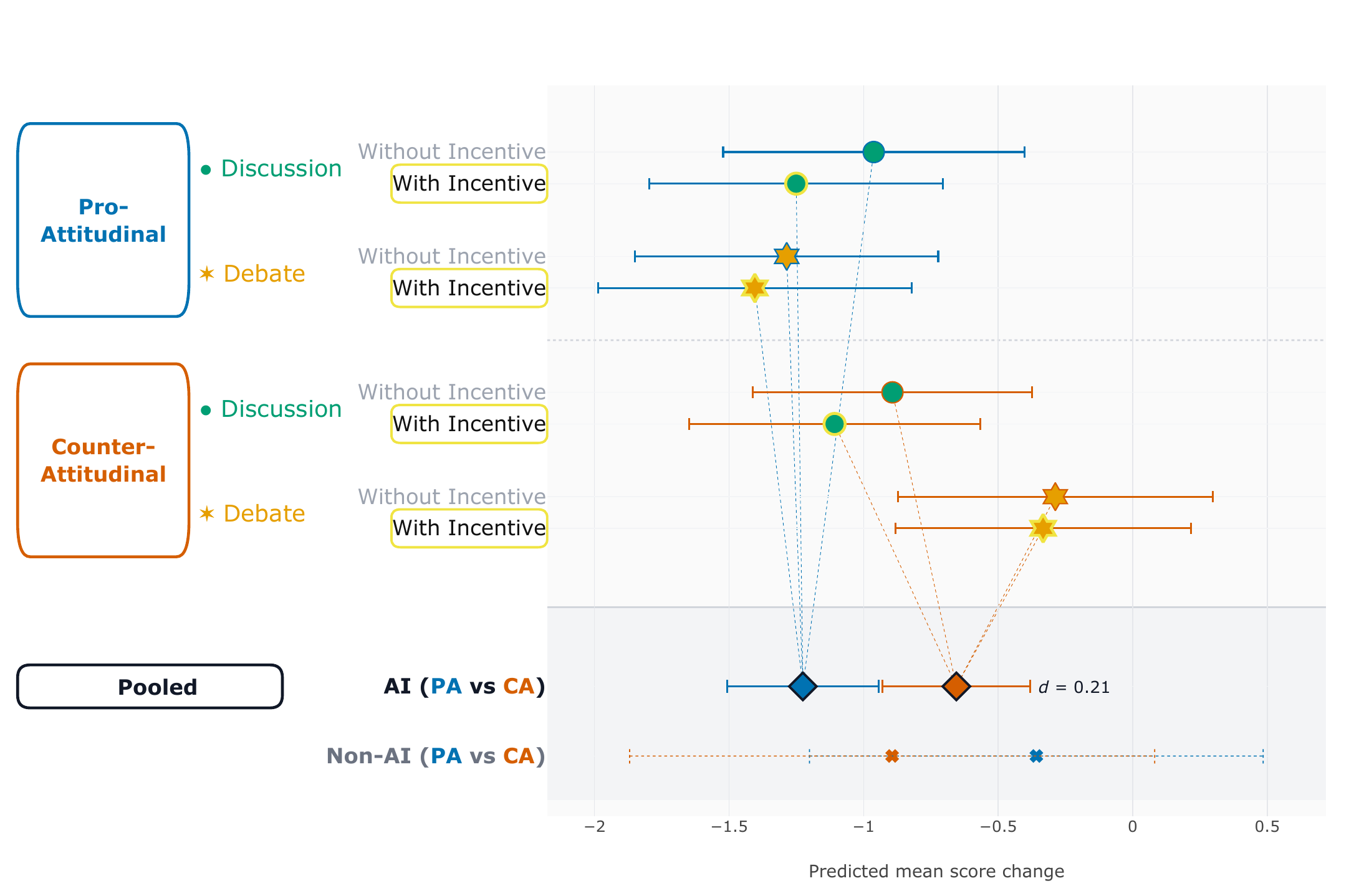}
\caption{Comparison of estimated marginal mean changes in affective differentiation score, where lower value indicates greater reduction in affective polarization. Counter-attitudinal advocacy against an AI partner resulted in weaker depolarization effect than pro-attitudinal advocacy in an equal-weight pooled comparison.}\label{fig_dotplot_affective_differentiation}
\end{figure*}

As a proxy for affective polarization, we define the outcome variable of ``affective differentiation" to capture the extent to which one's opinions of co-partisans and out-partisans differ. 
Participants in the pro-attitudinal AI conditions showed greater pre-post reduction in affective differentiation than those who engaged in a counter-attitudinal dialogue ($p=0.008, d=0.21$; Figure \ref{fig_dotplot_affective_differentiation}), contrary to our hypothesis H1b, which predicted counter-attitudinal superiority. When the instruction was pro-attitudinal advocacy---writing arguments that were congruent to their existing attitudes---the presence of an AI dialogue partner marginally enhanced the reduction in affective differentiation compared to the non-AI reference task ($p=0.058, d=0.33$). When counter-attitudinal advocacy was required, the non-AI reference task led to a modest reduction in affective differentiation and the presence of an AI partner did not directionally improve the outcome ($p=0.681, d=0.08$). As a result, our hypothesis that conversing with an AI partner on a political issue would reduce affective polarization across the board (H1a) was unsupported in a pooled comparison between the AI and non-AI arms. 

In absolute terms, a depolarization intervention in the form of an attitude-congruent discussion or debate with an AI partner could reduce affective differentiation relative to the pre-treatment level,  with a mean reduction equivalent to $0.27 \, SD$ ($\mathrm{CI}\,[0.09, 0.46]$) of all pre-treatment measurements for discussions and $0.34 \, SD$ ($\mathrm{CI}\,[0.15, 0.50]$) for debates. 

To understand argumentative dialogues' effect on opinion polarization, we compared the interlocutors' stance on the topic of discourse before and after the dialogue, and treated movement towards the opposite side as depolarization. Pro-attitudinal AI dialogues similarly resulted in stronger opinion shift towards the opposite side than counter-attitudinal ones ($p=0.021, d=0.20$). The presence of an AI partner significantly increased the stance movement relative to the non-AI task in a pro-attitudinal setup ($p=0.017, d=0.43$), whereas an AI partner did not shift participants' stances more strongly than the non-AI task when counter-attitudinal advocacy was required.

\subsection*{Dichotomous cognitive empathy pathways across argument side}

To assess participants' propensity to adopt an alternative psychological point of view, both in response to the dialogue and as a general disposition, we separately included scales for measuring cognitive empathy at state \cite{shen_scale_2010} and trait \cite{davis_measuring_1983} levels. State empathy, which is more strongly influenced by mental processes during an intervention, allowed for between-conditions comparisons after controlling for participants' pre-treatment trait empathy. Trait empathy, which reflects one's general disposition, was used to capture changes over the entire course of the two-session study.

The presence of an AI partner had disparate effects on cognitive state empathy across argument sides. We hypothesized that counter-attitudinal advocacy against an AI partner would result in higher state empathy than pro-attitudinal advocacy (H2b), but the opposite was marginally supported ($p=0.053$). Whereas a dialogue with an AI partner made participants significantly more empathetic than in a non-AI task in a pro-attitudinal setup ($p<0.001, d=0.74$), the effect was reversed in a counter-attitudinal setup, where an AI partner would elicit significantly lower state empathy than a corresponding non-AI task ($p=0.024, d=0.53$). Consequently, our hypothesis that AI presence enhanced cognitive state empathy (H2a) was unsupported in an equal-weight pooled comparison.

\begin{figure*}[t]
\centering
\includegraphics[width=0.9\textwidth]{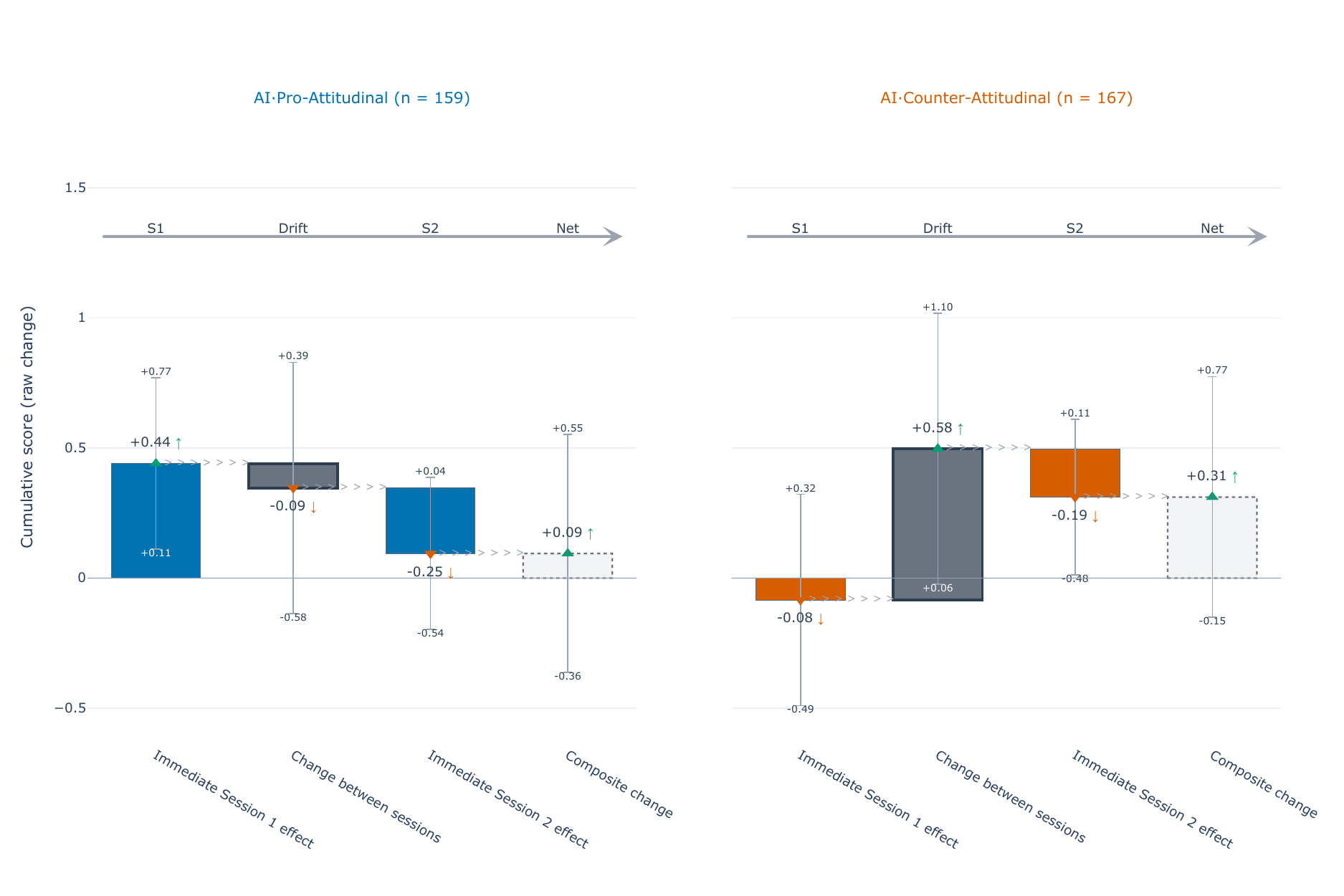}
\caption{A waterfall chart showing the changes in cognitive trait empathy throughout the study. The second and third bar start where the previous one ends, showing cumulative changes at that point.}
\label{fig_waterfall_trait_empathy}
\end{figure*}

In order to observe the long-term impact of AI-based argumentation exercises on cognitive empathy, the study was designed with two sessions, allowing for measurements at four time points. We examined the within-session changes (``S1" and ``S2"), the change observed between the end of the first session and the beginning of the second session (``Drift"), and the end-to-end change from the two-session treatment (``Net"). As shown in Figure \ref{fig_waterfall_trait_empathy}, neither the pro-attitudinal nor counter-attitudinal conditions for the AI task resulted in a statistically significant end-to-end change in cognitive trait empathy, but the two argument sides followed distinct pathways that saw statistically significant changes at different periods. Participants in the counter-attitudinal AI conditions recorded an increase in cognitive trait empathy during the gap between two sessions, both in absolute terms ($p=0.029, d_z=0.17$, or 0.58-point increases in IRI-PT \cite{davis_measuring_1983}) and compared to the non-AI reference ($p=0.031; n=10$), despite having recorded no significant change within either session. The pre-treatment trait empathy of Session 2 was also significantly higher than the pre-treatment level of Session 1 for participants assigned to the counter-attitudinal dialogues ($p=0.041, d_z=0.16$). The pro-attitudinal assignment significantly increased trait empathy within Session 1 ($p=0.009, d_z=0.21$, or a 0.44-point increase in IRI-PT) but produced no further boost in Session 2 or between sessions.
These observations, however, came from hypothesis-generating analyses that fell outside the scope of the preregistration.

\subsection*{AI partner's collaborative conversation style improves depolarization effects in counter-attitudinal dialogues}

We expected the dialogue mode of discussion to result in more favorable depolarization outcomes in both pro- and counter-attitudinal conditions. Informed by studies on classroom-based debates \cite{budesheim_consider_1999, de_conti_debate_2013} and competition dynamics \cite{klimecki_impact_2020}, we hypothesized that the more adversarial debate framing would ``backfire" when participants performed pro-attitudinal advocacy, motivating interlocutors to fend off challenges to their pre-existing opinions and negatively perceive the opponent's intention (H4a). There was no significant difference across dialogue mode for the pro-attitudinal dialogues to support H4a. 
For counter-attitudinal advocacy, we hypothesized that the cooperative context of a discussion would create a sense of emotional safety that increases the reduction in affective differentiation (H4b). Our data supported H4b, as participants in counter-attitudinal dialogues recorded greater reduction in affective differentiation when they engaged in a discussion as opposed to a debate ($p=0.021, d=0.26$). Similarly, the mode of discussion yielded less undesirable outcomes than debate in state empathy ($p=0.048, d=0.27$), trait empathy ($p=0.016, d=0.25$), and support for democratic norms ($p=0.010, d=0.26$) in counter-attitudinal dialogues.

\subsection*{Effects of financial incentives on counter-attitudinal advocacy}
We anticipated that the presence of an additional financial incentive would encourage participants to exert more effort in the challenging task of counter-attitudinal advocacy, leading to improved outcomes. While the offer of a bonus payment did not reduce affective differentiation after counter-attitudinal dialogues (contrary to H4c), participants in counter-attitudinal conditions reported higher state empathy ($p=0.044, d=0.28$) and marginally higher critical reflection ($p=0.057, d=0.40$) when the offer was present. Financial incentives also resulted in higher critical reflection among participants in the debate conditions ($p=0.002, d=0.70$), across both argument sides.

\subsection*{Muted effects on support for democratic norms and political intolerance}

With respect to the exploratory hypotheses that the presence of an AI argumentative partner would enhance support for democratic norms (H3a) and reduce political intolerance (H3b) relative to the non-AI task, the immediate pre-post changes failed to establish the benefits of AI presence. Pro-attitudinal advocacy AI dialogues significantly reduced the score for support for democratic norms relative to the non-AI task ($p=0.024, d=0.35$), which contributed to a directionally worse outcome for the AI arm than the non-AI arm in a pooled comparison, refuting our hypothesis of AI benefit (H3a). There was little difference in pre-post changes to political intolerance between a non-AI task and AI dialogues in the corresponding argument side to support our hypothesis that an AI dialogue could make interlocutors more tolerant (H3b). While these results are in agreement with prior work that decouple improved opinions of out-partisans from reduced anti-democratic attitudes \cite{voelkel_interventions_2022, druckman_correcting_2023}, it is worth noting that the two outcome variables suffer from ceiling and floor effects, with 44\% of pre-treatment responses reporting the highest possible score for support for democratic norms and 40\% reporting the lowest possible score for political intolerance. The ceiling and floor meant our measures had greater sensitivity to worsening than to improvement, which likely underestimated changes in the beneficial direction.

Over the course of the entire study, the participants who completed both sessions in a counter-attitudinal AI condition recorded a positive end-to-end net increase in support for democratic norms ($p=0.014, d_z=0.19$), while the pro-attitudinal dialogues yielded no significant end-to-end change. 
Political intolerance was reduced by both pro- ($p=0.013, d_z=0.20$) and counter-attitudinal ($p=0.021, d_z=0.18$) AI dialogues in Session 1 in absolute terms, yet the reduction reverted completely by the start of Session 2.

\section*{Discussion}\label{sec12}

In our results, the most robust pattern was heterogeneity of effects across the AI conditions: attitude-congruent dialogues produced larger immediate changes in affective differentiation and opinion polarization than attitude-incongruent dialogues, whereas the pooled comparisons between the AI and non-AI arms were more limited by design asymmetry and statistical power. 

Our study shows the potential of attitude-congruent human-AI argumentative dialogues in reducing both affective and opinion polarization, both in absolute terms and relative to an essay writing task without an AI partner ($d=0.33$ for affective polarization and $d=0.43$ for opinion polarization). While modest by conventional statistical benchmarks, the effect sizes from the short exercise are comparable to what other studies have reported for the difficult task reducing partisan animosity \cite{voelkel_megastudy_2024, combs_reducing_2023, holliday_why_2025} and changing attitudes \cite{costello_durably_2024}.

Contrary to results from debates between human interlocutors in a classroom setting \cite{budesheim_consider_1999, de_conti_debate_2013, lord_biased_1979}, pro-attitudinal advocacy did not appear to result in entrenchment in the presence of an AI partner, and those who performed counter-attitudinal advocacy showed weaker immediate depolarization effects and lower cognitive state empathy. Similarly, the participants in the counter-attitudinal AI dialogues---whom we expected to become more open-minded from utilizing the cover of devil's advocacy to explore ideas that might otherwise be socially controversial \cite{warren_navigating_2019}---reported significantly weaker cognitive state empathy than those in the non-AI reference task. 
It appeared that counter-attitudinal participants, upon seeing an apparently knowledgeable AI chatbot eloquently backing their true opinions, were discouraged from reconsidering their position in the same way one might do in a solitary task.
The departure of our results from prior literature suggests that an argumentative dialogue may influence interlocutors' attitudes via different mechanisms in human-AI interaction than in the human-human setting emphasized in earlier work.

The study was designed to test the depolarization potential of a dialogue intended to elicit perspective-taking and self-persuasion, specifically within the setup of counter-attitudinal advocacy. Counter-attitudinal advocacy was not as effective as hypothesized; instead, the results support the explanation that depolarization was partly driven by the engagement with alternative perspectives---albeit through exposure to the AI partner's arguments rather than the self-generated ones---and persuasion. In line with existing literature on AI's persuasiveness in political messaging \cite{salvi_conversational_2025, argyle_testing_2025, hackenburg_levers_2025, goldstein_how_2024, bai_llm-generated_2025, lin_persuading_2025} and human's openness to opposing perspectives that originate from AI \cite{lu_how_2025}, participants in the pro-attitudinal advocacy conditions---who processed counter-attitudinal messages from the AI partner, regardless of the dialogue mode---softened their stance on the issue discussed and perceived people who hold opposing opinions more favorably. Our mediation analysis found that 15.0\% of the pro-attitudinal dialogues' outperformance in reducing affective differentiation is consistent with a pathway through stance change, strengthening the explanation while highlighting the role of opinion-independent mechanisms, such as receptiveness \cite{santoro_listen_2025, yeomans_conversational_2020} and correction of misperception built on stereotypes \cite{druckman_misestimating_2022, pasek_misperceptions_2022}.

While the pro-attitudinal AI dialogues were effective in shifting the interlocutors' stance on the discourse topic towards the other side, no spillover change was observed for the three other issues not addressed in the dialogue. Moreover, the pro-attitudinal participants' level of affective differentiation at the start of Session 2, which was anchored on a new discourse topic for the session, did not retain the reduction recorded in Session 1. Although our experimental setup did not allow us to assess the durability of affective change with respect to opponents on a particular issue, the significant rebound of affective differentiation between sessions shows the limited long-term effect that a single session may have in reducing animosity towards ideological opponents. 

The pro-attitudinal AI dialogues outperformed counter-attitudinal ones in eliciting cognitive state empathy during the sessions, but the higher state empathy did not translate to a sustained increase in trait empathy. While trait empathy initially increased after pro-attitudinal dialogues in Session 1, the pro-attitudinal dialogue participants who returned for Session 2 did not continue to build trait empathy in Session 2 or between sessions. The counter-attitudinal dialogues, in contrast, failed to boost trait empathy during Session 1 or Session 2 but produced a sizable increase between sessions (Figure \ref{fig_waterfall_trait_empathy}). 
The contrast suggests that self-generated counter-attitudinal arguments may be more challenging to make and more uncomfortable to accept, but as the initial resistance from the exchange subsides, interlocutors may later ruminate on the arguments as they work to resolve the uncertainty about what they truly believe in \cite{higgins_self-discrepancy_1989}.

Existing theories on self-generated arguments \cite{lemmen_convince_2020, baldwin_examining_2013, higgins_self-discrepancy_1989, brinol_self-generated_2012, miller_counter-attitudinal_2001} and sleeper effect \cite{kumkale_sleeper_2004} in persuasion offer further explanations for the delayed trait empathy boost from the counter-attitudinal dialogues. 
Studies show that whereas externally-provided arguments for beneficial behaviors are easier to be understood, self-generated arguments create intrinsic motivation to process the ideas more elaborately, leading to higher perceived importance of the behaviors \cite{lemmen_convince_2020, baldwin_examining_2013}. The pro-attitudinal dialogue participants had the counterarguments clearly laid out by the AI partner over multiple rounds of exchange, and the effortless understanding of alternative perspectives may have led to the instant empathy boost. For participants who performed counter-attitudinal advocacy, the eloquent AI arguments that reinforced their true position may have functioned---in the framing of sleeper effect \cite{kumkale_sleeper_2004}---as a discounting cue that suppressed an attitude change in the moment but was forgotten more quickly than the self-generated arguments.

The study has several limitations affecting its generalizability. Although we have targeted a diverse demographic composition matching the latest US census data, the choice of Prolific as our recruitment platform predetermined a participant pool that is likely more digitally savvy and less financially stable than the overall population. 
Prolific's default payment structure of awarding a fixed amount for a task means participants on the platform are generally financially incentivized to speed through a survey rather than perfecting each response. The substantial proportion (27\%) of complete submissions removed from Session 1 after quality checks underscores a general lack of attention and effort among online participants, and it remains an open question whether the study participants engaged with the AI chatbot at a depth sufficient to uncover the full effects of a human-AI dialogue. On the flip side, however, the same removal rate can be interpreted as evidence of a robust multi-layer quality-control mechanism that safeguarded our statistical results from inattentive and automated survey responses.

As a behavioral compliance check for the argument side manipulation, we classified participant responses using LLM-based stance detection and discovered that 17.3\% of participants in the counter-attitudinal conditions failed to argue against the AI partner (see Supplementary Section 5.1). We proceeded the analysis on an intention-to-treat principle, which reflects the treatment effects in a realistic setting. The low-compliance rate likely biased the results towards the null, which was a problem that could not be resolved post hoc---a simple removal would limit our sample to participants who were open-minded and comfortable playing devil's advocate. A summary of per-protocol analysis, which shows a similar pattern albeit with reduced power, is provided in Supplementary Section 5.3.

Our own payment structure introduces additional complications to our assessment of human-AI dialogue as a scalable intervention. On one hand, the financial motivation was inseparable from our participant recruitment method, which prevented us from answering the question of how much mental effort interlocutors might voluntarily put into a dialogue with an AI partner had they participated purely for sport or self-improvement. On the other hand, since we targeted a high hourly rate of \$24 (or £18, as the pound sterling is the default currency on Prolific) to compensate for the high mental demand, it may have intensified the extrinsic financial motivation \cite{murayama_neural_2010, vlaev_changing_2019} and diminished the relative value of our additional performance-based incentive, leading to the inclusion of participants who may be reluctant to engage in a dialogue and the muted effect of a financial incentive as an experimental variable.

Our sampling strategy prioritized detecting differences between AI conditions, rather than comparing the AI and non-AI arms\footnote{The non-AI reference conditions required considerable mental effort and should not be understood as a trivial placebo. The mean cognitive load scores for the AI and non-AI arms were 7.02 and 6.41, respectively, on a 1-9 scale.}. In the face of lower-than-expected effect sizes, the setup successfully captured the difference across argument sides, our foremost experimental variable, but was underpowered to make strong claims about the intergroup benefits stemming from the presence of an AI partner. The low survey submission count (under 10\% of the pool of qualified Prolific participants), substantial exclusion rate due to quality issues (24\% across two sessions), and high attrition between sessions (34\%) further constrained the achieved sample. The between-sessions attrition rate varied across groups, with women and Republicans being more likely to drop out after Session 1 (Supplementary Section 5.2). As a result of the small sample size, responses from the non-AI arm could not establish a confident baseline for longitudinal analysis.

Lastly, our non-AI reference task was solitary rather than interactive. Consequently, our contrast between the AI and non-AI conditions could not isolate the effect of the agent (an AI partner) from that of the medium (a multi-turn dialogue). Future studies could introduce real-time human-human dialogues as a causal control condition to better isolate the effects.

Prior studies have presented promising evidence that AI-generated text can positively influence political attitudes \cite{costello_durably_2024, salvi_conversational_2025, sevi_ai_2026} and reduce polarization \cite{hruschka_reducing_2026, walter_using_2025}. Our results show that the effects on intergroup attitudes are heterogeneous across dialogue design, chiefly the specification of argument side, implying that the presence of AI alone is not sufficient. Attitude-congruent dialogues appear to achieve immediate softening through the mechanisms of persuasion and conversational receptiveness, whereas our exploratory longitudinal evidence suggests that self-generated counter-attitudinal arguments may produce delayed but more durable perspective-taking gains, drawing a parallel with the theory of sleeper effects, although this pattern requires validation with a larger longitudinal sample. Designing human-AI dialogues for attitude change therefore likely requires matching dialogue design with intended outcomes and time horizon, a judgment call that would benefit from further empirical research into the underlying mechanisms.

\section*{Methods}\label{sec11}

US-based online participants identifying as either a Republican or Democrat in the profile completed up to two sessions (N=527 for Session 1, of which 350 also completed Session 2) of a written argumentation task in a customized Qualtrics\footnote{\url{https://www.qualtrics.com}} survey (see Supplementary Sections 1 and 2 for details about participant recruitment, quality checks and interaction design). 
In each session, participants first answered questions on demographics, technology use and political attitudes, before being presented with pre-post questionnaires and either an AI chatbot or a writing prompt for the non-AI task. The median AI dialogue duration was 16.3 minutes.

The study was preregistered on the Open Science Framework\footnote{\url{https://osf.io/cgxuj/overview?view_only=785731015d6e4b1cb75dfab77d2a2624}} prior to data collection. Recruitment and payment of the participants took place on Prolific Academic\footnote{\url{https://www.prolific.com}}, informed by prior studies involving online crowdworkers \cite{palan_prolificacsubject_2018, peer_beyond_2017, zhu_learn_2025, douglas_data_2023}. Voluntary consent for participation was collected from all participants at the start of each session. Personally identifiable information was algorithmically scrubbed with Presidio\footnote{\url{https://microsoft.github.io/presidio}} before user input was sent to OpenAI's server. To minimize demand effects, the participants were not initially informed about the purpose of the study, and the task was described as data collection. The survey ostensibly invited them to contribute diverse opinions on political issues for AI training, and displayed a chatbot-based ``role-playing game" to assign them into either pro- or counter-attitudinal argument side. At the conclusion of data collection, a debrief message was sent to participants as a Prolific on-site message to explain the research goal of exploring AI-based depolarization strategies. 

The chatbot was powered by OpenAI's GPT-5.1 model\footnote{\url{https://openai.com/index/gpt-5-1}}, configured with custom prompts and rule-based progression logic. Before entering the chat interface, participants were introduced to our ``role-playing game" and asked to imagine themselves at a party where an outspoken acquaintance makes their opinions known. The instruction then invited participants to take the opposite position and become a sparring partner. As soon as participants entered the chat interface, the AI partner would send a provocative remark stating its position and challenge participants to come up with counterarguments. Depending on the exact experimental condition, the AI would take either an adversarial (``No offense, but...") or collaborative (``Help me see what I'm missing...") tone, and may offer £1 for changing its mind (in a debate) or showing its blind spots (in a discussion). Afterwards, participants were guided through a sequence of three steps---presenting an argument, responding to the opponent, wrapping up---for up to three repetitions on different sub-topics. Short or vague input from participants would be met with a request for elaboration, and repeated disengagement would lead to an early conclusion. In a debate, the AI partner would persistently try to win the argument. In a discussion, the AI partner would try to find a common ground without necessarily compromising. At the conclusion of the dialogue, the AI partner would comment on the merits of participants' arguments and provide a reflection prompt.

With a three-tier strategy, our preregistration identified affective differentiation (pre-post change in the difference between respondents' evaluation of their own group and the opposing group, based on the stereotype content model \cite{fiske_stereotype_2018, schieferdecker_affective_2024}) as the primary outcome of interest, and cognitive state empathy \cite{shen_scale_2010} (post-treatment only) as the secondary outcome. Support for democratic norms and political intolerance \cite{bos_are_2023, sullivan_sources_1981} (both pre-post changes) were included with exploratory hypotheses given their conceptual relevance to intergroup relations, notwithstanding the anticipated measurement difficulties arising from ceiling and floor effects. Additional exploratory variables included cognitive trait empathy (pre-post change; the perspective-taking subscale of the Interpersonal Reactivity Index \cite{davis_measuring_1983}), stance change toward the opposing position (pre-post change), and critical reflection \cite{kember_development_2000} (post-treatment only). As further discussed in Supplementary Section 2.4, the survey scales with at least four items all achieved high internal consistency scores, which suggests our participants responded meaningfully to our questions. Our quality checks slightly shifted the final sample's internal consistency scores in directions that are consistent with the removal of inattentive straight-line response patterns.

We further defined a hierarchical relationship between the experimental variables (visualized in the tree diagram in Supplementary Figure S1). The argument side assignment, which determined how the participants engaged with their pre-existing opinions, represented the primary experimental manipulation within the AI arm. The dialogue mode, which specified the AI partner's behavior, was expected to influence the ease with which the participants expressed arguments for the assigned side. The financial incentive, designed to motivate participants without altering the task specification for either the participants or the AI partners, was included to produce a subtle boost for the specific combinations of the two other variables. The non-AI conditions lived on a separate branch and provided an effort-matched reference for an argumentative task for pooled comparisons, although they did not isolate the effect of having an AI dialogue partner from that of participating in an interactive dialogue.

To perform hypothesis testing, we fitted two analogous mixed-effects models for each outcome variable to separately test hypotheses about the overall effects of AI interaction and the roles of experimental variables within the AI conditions (see Supplementary Section 3 for details). For comparison between AI and non-AI conditions, we performed equal-weight pooling across all eight AI conditions and performed Wald tests on the contrasts. Similarly, to test hypotheses relating to argument side, participants were pooled equally across dialogue mode and financial incentive. Between-condition effect sizes were reported as Cohen's $d$, calculated as the Wald contrast estimate divided by the model residual standard deviation; longitudinal changes within the same condition were reported as Cohen's $d_z$, which is the mean change divided by the standard deviation of changes. We reported unadjusted p-values and treated exploratory findings as hypothesis-generating.

\backmatter

\bmhead{Supplementary information}
The research plan was preregistered on the Open Science Framework (\url{https://osf.io/cgxuj/overview?view_only=785731015d6e4b1cb75dfab77d2a2624}) prior to data collection. A supplementary material document is included to provide detailed explanations about the study. The accompanying dataset and code can be downloaded from Harvard Dataverse (\url{https://doi.org/10.7910/DVN/KPTFKB}).

\section*{Declarations}
The study was conducted remotely by researchers based at Saarland University in Germany. Participants located in the United States were recruited through an online participant-recruitment platform; no US-based research institution or physical research site was involved. The research was performed in accordance with the Code of Ethics and Professional Conduct of the Association for Computing Machinery, following a study protocol approved by the Ethical Review Board of the Faculty of Mathematics and Computer Science at Saarland University (\#25-08-12). All participants provided voluntary informed consent after receiving detailed information about the study procedures; information about the specific research aims was withheld until after participation to minimize demand effects, after which participants were fully debriefed.

The project has been financially supported by the Alexander von Humboldt Foundation, its founder the German Federal Ministry of Education and Research (BMBF), and the German Academic Exchange Service (DAAD). Open access to this manuscript is supported by the German Research Foundation (DFG). The funders had no influence in the design, implementation or publication of the study.

\section*{Author contributions}
J.Z., S.N., U.N. and I.W. designed research; J.Z. and S.N. created tools and performed research; J.Z. wrote the paper; C.Z., U.N. and I.W. revised the paper. J.Z. is the corresponding author.

\bibliography{references}%

\end{document}